\documentclass[pre,amsmath, twocolumn,showpacs]{revtex4-1}
\usepackage{graphicx}
\usepackage{color}
\usepackage{latexsym}

\begin{document}

\title{Information-theoretic bound on the entropy production to maintain a classical nonequilibrium distribution using ancillary control}

\author{Jordan M. Horowitz}
\author{Jeremy L. England}
\affiliation{Physics of Living Systems Group, Department of Physics, Massachusetts Institute of Technology, 400 Technology Square, Cambridge, MA 02139}

\date{\today}

\begin{abstract}  
There are many functional contexts where it is desirable to maintain a mesoscopic system in a nonequilibrium state.  However, such control requires an inherent energy dissipation.  In this article, we unify and extend a number of works on the minimum energetic cost to maintain a mesoscopic system in a prescribed nonequilibrium distribution using ancillary control.  For a variety of control mechanisms, we find that the minimum amount of energy dissipation necessary can be cast as an information-theoretic measure of distinguishability between the target nonequilibrium state and the underlying equilibrium distribution.  
This work offers quantitative insight into the intuitive idea that more energy is needed to maintain a system farther from equilibrium. 
\end{abstract}


\maketitle 

\section{Introduction}
Small systems are continually bombarded by noise from their surroundings.
Sometimes this noise is helpful; thermal and chemical fluctuations are the fuel that power biological molecular motors~\cite{Parrondo2002}. 
More often, though, noise is a nuisance.
Fluctuations in gene expression or transcription can lead to errors in downstream macromolecules, like RNA, that can be detrimental to a cell's function~\cite{Bialek}.
Noise can also interfere with the functioning of artificial mesoscopic devices, such as micromechanical~\cite{Li2011} and nanomechanical resonators~\cite{Tian09,Palomaki13}.
In all these situations, an ancillary control mechanism can be employed to suppress fluctuations.
This can take the form of a kinetic proofreading scheme~\cite{Bialek} or the addition of an auxiliary control device that employs active feedback, as in a Maxwell's demon~\cite{Horowitz2013,Munakata2012,Parrondo2015,Sandberg2014,Horowitz2014,Horowitz2014b,Shiraishi2014}.

No matter the control mechanism, the effect is to force the system into a statistical state distinct from its noisy equilibrium, where it will inevitably dissipate energy.
Thus maintaining a system away from equilibrium comes with an energetic cost.
Attempts at predicting the properties of such nonequilibrium states  by minimizing the energy dissipation have a long history, starting with Prigogine and coworkers~\cite{Kondepudi} within linear irreversible thermodynamics (see also~\cite{Maes2007}).
However, it seems no such general variational principle exists beyond the linear regime~\cite{Maes2007,Bruers2007,Poletinni2013}.
As such, our goal in this work is not to characterize the nonequilibrium state through a thermodynamic variational principle.
Instead, we aim to characterize the energetic requirement to hold an originally equilibrium system in a prescribed out-of-equilibrium state using an additional external control system that does not alter the original system's properties.  Indeed, previously in Refs.~\cite{Horowitz2017, Horowitz2015b}, we showed that for specific classes of externally imposed controls, this minimum energetic cost could be expressed simply in terms of the systems underlying equilibrium dynamics.
In this Article, we expand this program to include new control mechanisms, and in the process offer a unifying perspective on these previous results.
In particular, we demonstrate that for various control mechanisms the minimum entropy production (or dissipation) to keep a mesoscopic system in a specified nonequilibrium distribution can be expressed as a time derivative of the relative entropy between the target distribution and the uncontrolled equilibrium Boltzmann distribution.
This information-theoretic characterization quantitatively characterizes the intuitive notion that the farther a system is from equilibrium the more energy must be dissipated to maintain it.

\section{Setup}
\begin{figure}[htb]
\centering
\includegraphics[scale=.25]{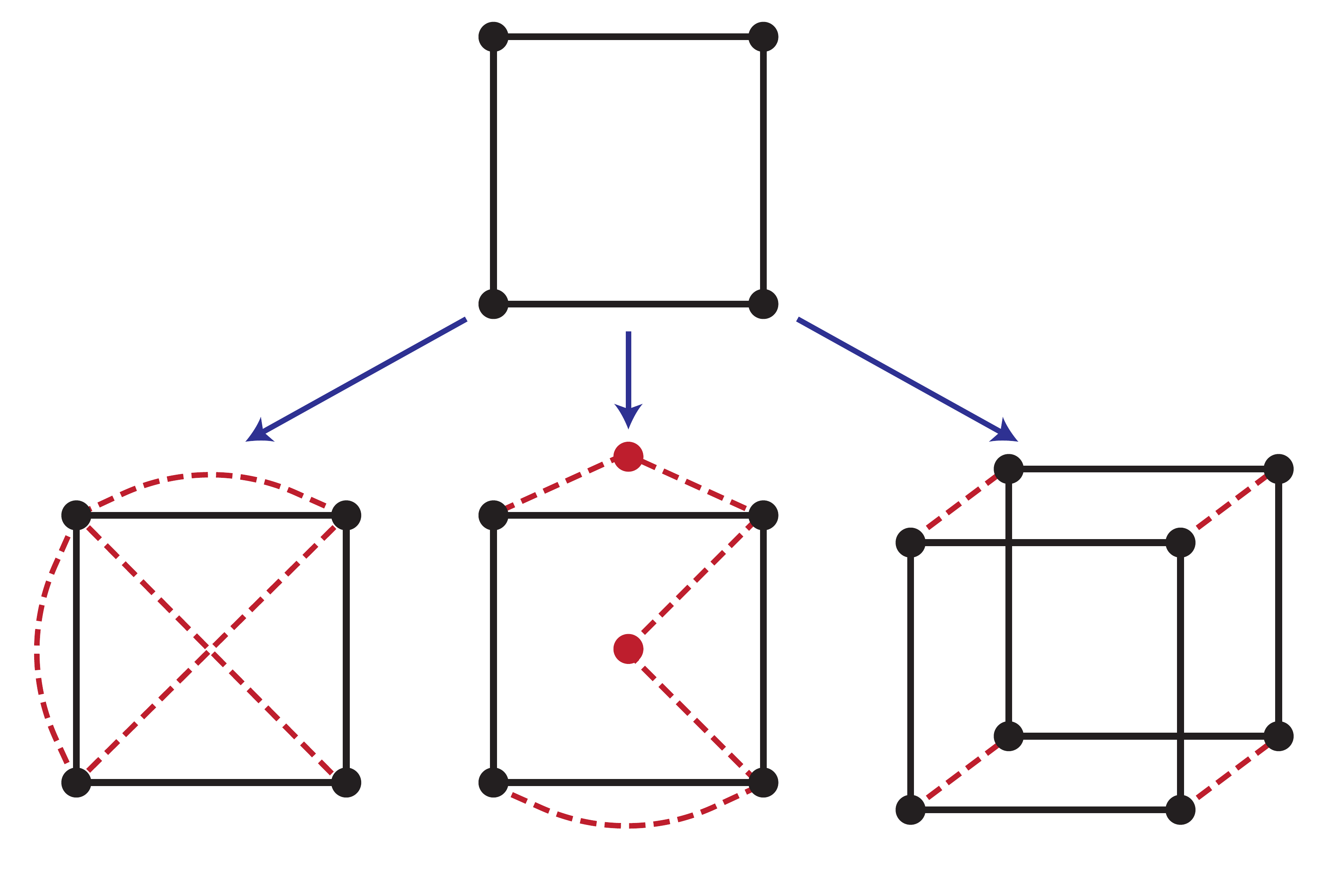}
\caption{Illustration of three types of control: (Top) Graph representation of the system's configuration space without control.  Mesoscopic configurations are represented as vertices (or nodes) with edges signifying allowed transitions.  (Bottom)  Control is implemented by adding additional edges (red dashed) or nodes (red dots) in order to drive the system into a nonequilibrium distribution.  From Left to Right: Edge control, Node control, and Auxiliary control.}
\label{fig:graph}
\end{figure}

We have in mind a small mesoscopic system making random transitions among a set of discrete mesostates, or configurations, $i=1,\dots, N$, each with (free) energy $E_i$.
We can visualize this dyanmics occurring on a graph (or network) like in Figure~\ref{fig:graph}, where each configuration is assigned a vertex (or node), and possible transitions are represented by edges (or links).

The dynamical evolution is modeled as a Markov jump process on our graph according to transition rates $W_{ij}$ from $j\to i$, with $W_{ij}\neq 0$ only when $W_{ji}\neq0$, so that every transition has a reverse.
As such, the system's time-dependent probability density $p_i(t)$ evolves according to the Master equation~\cite{VanKampen}

\begin{equation}\label{eq:master}
\partial_tp_i(t) = \sum_{j\neq i} W_{ij}p_j(t)-W_{ji}p_i(t),
\end{equation}

\noindent which constitutes a probability conservation equation with probability currents

\begin{equation}
J_{ij}(p)=W_{ij}p_j(t)-W_{ji}p_i(t).
\end{equation}

Now, in the absence of any external control, we assume that our system relaxes to a thermal equilibrium steady state at inverse temperature $\beta=1/k_{\rm B}T$, given by the Boltzmann distribution $p_i^{\rm eq}=e^{\beta(F^{\rm eq}-E_i)}$ with equilibrium free energy $F^{\rm eq}=-k_{\rm B}T\ln\sum_i e^{-\beta E_i}$.
To guarantee this, we impose detailed balance on the transition rates~\cite{VanKampen},

\begin{equation}\label{eq:db}
W_{ij}p_j^{\rm eq}=W_{ji}p_i^{\rm eq}.
\end{equation}

\noindent In equilibrium, each transition is counter-balanced by its reverse.
Our goal is then to maintain the system in a predetermined target nonequilibrium steady state $p^*\neq p^{\rm eq}$, and to characterize the minimum dissipation necessary.

\section{Minimum dissipation to suppress fluctuations} 

We are interested in pushing and holding our system into a statistical state distinct from the equilibrium Boltzmann distribution.
One could imagine a variety of schemes to accomplish this goal.
Perhaps the simplest, is if we have complete control to vary the system's energy function $\{E_i\}$.
We could then hold the system in the statistical state $p^*$ by shifting all the energy levels to $E^*_i=-k_{\rm B}T\ln p^*_i$, thereby making the target state the new equilibrium state: $p^*_i=e^{-\beta E^*}$.
After an initial transient relaxation, the system would then remain in $p^*$ indefinitely as it is in equilibrium.
While there is a one time energetic cost to vary the energy levels -- equal to the free energy difference~\cite{Esposito2011} --, the system can be held in $p^*$ for free.
Implementing such a protocol, however, generically requires very fine control over all the individual energies, which often is prohibitive~\cite{Esposito2011}.
As a result, there are a number of situations where this is not possible or not desirable.
For example, nature does not utilize this control mechanism; in cells, where the free energies of molecules are fixed, noise reduction is implemented by coupling together various driven chemical reactions that constantly burn energy~\cite{Bialek,Sartori2014,Sartori2015}.
Whereas fluctuations in quantum mesoscopic devices are often suppressed by coupling an auxiliary device that continually and coherently extracts noise through feedback~\cite{Tian09,Horowitz2013c, Hamerly2012,Schliesser08}.

Motivated by this observation, we analyze control mechanisms where we cannot alter the internal energies $\{E_i\}$.
Instead, the statistical state of our system is manipulated by introducing additional pathways.
In particular, the scenarios we address, depicted in Figure~\ref{fig:graph}, are : (i) Edge control -- additional driven transitions (or edges) are added with transition rates $\{M_{lk}\}$, which model the coupling of additional thermodynamic reservoirs;  (ii) Node control -- additional configurations are incorporated and coupled to the original network through driven transitions with rates $\{M_{lk}\}$, allowing for ancillary intermediate configurations, such as in dissipative catalysis;  (iii) Auxiliary control -- an entirely new system is coupled to the controlled system, as in feedback control; and finally (iv) Chemical control -- where new chemical reactions are included.  Though ostensibly a special case of edge control, it adds new complications due to the possibility of breaking conservation laws.
The addition of such controllers alters the system's dynamics leading to a modified Master equation (cf.~\ref{eq:master})

\begin{align}\label{eq:master2}
\partial_tp_i(t) &= \sum_{j\neq i} W_{ij}p_j(t)-W_{ji}p_i(t)+\sum_{j\neq i} M_{ij}p_j(t)-M_{ji}p_i(t)\\
&=\sum_{j\neq i}J_{ij}(p)+J^M_{ij}(p).
\end{align}
Yet no matter which control mechanism is employed, we assume the net effect is to push our target system into the nonequilibrium steady state $p^*$.
While designing such control is generically a challenging problem, we take it as a given and instead focus on the minimum cost.

Our only assumption is that the additional control transition rates satisfy a local detailed balance relation connecting them to the entropy flow into the thermodynamic reservoir that mediates the transition~\cite{Esposito2010b,Seifert2012}:

\begin{equation}\label{eq:ldb}
\ln \frac{M_{kl}}{M_{lk}}=\Delta s^{\rm e}_{kl}.
\end{equation}
For example, if we implement control by coupling a thermal reservoir at a different inverse temperature $\beta^\prime$, then we require $\ln(M_{kl}/M_{lk})=\beta^\prime(E_l-E_k)$, which is proportional to the heat flow into the environment.

The local detailed balance relation implies that our super-system, composed of the system of interest and the controller, with rates $\{{\mathcal W}^\kappa_{ij}\}=\{W_{ij},M_{ij}\}$, where $\kappa$ specifies an uncontrolled or controlled transition, satisfies the second law of thermodynamics: Namely, the (irreversible) entropy production is positive

\begin{equation}\label{eq:entProd}
{\dot S}_{\rm i}=k_{\rm B}\sum_{i>j,\kappa}J^\kappa_{ij}(p)\ln\frac{{\mathcal W}^\kappa_{ij}p_j}{{\mathcal W}^\kappa_{ji}p_i}\ge 0, 
\end{equation}
which is typically split into the derivative of the Shannon entropy

\begin{equation}
\partial_tS=-k_{\rm B}\partial_t\sum_i p_i\ln p_i=k_{\rm B}\sum_{i>j,\kappa}J^\kappa_{ij}(p)\ln\frac{p_j}{p_i},
\end{equation}
and the entropy flow

\begin{equation}
{\dot S}_{\rm e}=k_{\rm B}\sum_{i>j,\alpha}J^\kappa_{ij}(p)\ln\frac{{\mathcal W}^\kappa_{ij}}{{\mathcal W}^\kappa_{ji}}.
\end{equation}
Our goal will be to bound the entropy production (or dissipation) over all controls that fix the steady state distribution to be $p^*$.
Due to the local detailed balance relation [\eqref{eq:db} and \eqref{eq:ldb}], we can always connect the dissipation to the underlying energetics.
Thus, our bound on the entropy production can always be reframed as a minimum energetic cost.

\subsection{Edge control} 
We begin our investigation with the edge control scheme, where we add a collection of additional edges to the graph, corresponding to new transitions mediated by additional thermal or chemical reservoirs.
This analysis was originally carried out in \cite{Horowitz2017}.
We briefly review it here, as this control scheme is the simplest and all the following developments will build on it.

In this scenario the super-system produces entropy in the controlled steady state $p^*$ at a rate

\begin{equation}
{\dot S}_{\rm i}=k_{\rm B}\sum_{i>j}J_{ij}(p^*)\ln \frac{W_{ij}p^*_j}{ W_{ji}p^*_i}+k_{\rm B}\sum_{k>l}J^M_{kl}(p^*)\ln \frac{ M_{kl}p^*_l}{M_{lk}p^*_k}.
\end{equation}
We now wish to bound this sum solely in terms of properties of the system's environment as codified by the $\{W_{ij}\}$ and the target distribution $p^*$.

To this end, we observe that not only is the total entropy production positive, but link by link the entropy production is positive, $J_{kl}\ln (M_{kl} p_{l}/M_{lk}p_{k})\ge 0$~\cite{Shiraishi2014}, which follows readily from the inequality $(x-y)\ln(x/y)\ge 0$.
Thus, each control edge only contributes additional dissipation, implying that the only unavoidable dissipation occurs along the system's original links:
\begin{equation}\label{eq:edgeBound1}
{\dot S}_{\rm i}\ge {\dot S}_{\rm min} =k_{\rm B}\sum_{i>j} J_{ij}(p^*)\ln \frac{W_{ij} p^*_{j}}{W_{ji}p^*_{i}}\ge 0.
\end{equation}
No matter how control is implemented, the system will inevitable make jumps along the original links, and those will on average dissipate free energy into the environment when the system is held in the target state $p^*$.

We now offer some physical insight into the meaning of \eqref{eq:edgeBound1}.
To this end, we recognize that ${\dot S}_{\rm min}$ is the entropy production rate of the equilibrium dynamics when the statistical state is the target state $p^*$.
In other words, it represents the instantaneous entropy production we would observe if we turned off the control and allowed $p^*$ to begin to relax to equilibrium.
An enlightening reformulation of this observation is offered by recalling the intimate connection between the time derivative of the relative entropy, $D(f||g)=\sum_i f_i\ln (f_i/g_i)$~\cite{Cover}, and the entropy production rate: 

\begin{equation}
-k_{\rm B}\partial_t D(p(t)||p^{\rm eq})=\sum_{i>j}J_{ij}(p)\ln\frac{W_{ij}p_j}{W_{ji}p_i},
\end{equation}
which is a direct consequence of detailed balance \eqref{eq:db}.
As such, we immediately recognize that the minimum dissipation \eqref{eq:edgeBound1} can be equivalently formulated as

\begin{equation}\label{eq:edgeBound2}
{\dot S}_{\rm i}\ge {\dot S}_{\rm min}=-k_{\rm B}\partial^{\rm eq}_t D(p^*||p^{\rm eq}),
\end{equation}
where the derivative $\partial_t^{\rm eq}$ should be understood to operate on $p^*$ as if it were evolving under the uncontrolled equilibrium dynamics.
As the relative entropy is an information-theoretic measure of distinguishability~\cite{Cover}, \eqref{eq:edgeBound2}  quantifies precisely the intuitive fact that it costs more to control a system the farther it is from equilibrium.

We note that this analysis immediately offers the condition under which we saturate the minimum.
As our bound originates in setting aside the extraneous entropy production due to the control transitions, we immediately find as a consequence that this additional entropy production is zero when the control transitions operate thermodynamically reversibly.
This requires them to operate much faster than the system dynamics, so that at any instant the system is locally detailed balanced with respect to the control transitions on a link-by-link basis.
In other words, the optimal transition rates must verify

\begin{equation}
M_{kl}^*p^*_l=M_{lk}^*p_k^*.
\end{equation}
Indeed, this implies the optimal dissipation on each driven link (cf.~\ref{eq:ldb}) should be 

\begin{equation}
\Delta s^*_{kl}=\ln \frac{M^*_{kl}}{M^*_{lk}}=\ln \frac{p^*_k}{p^*_l}.
\end{equation}

\subsection{Node control} More than a fundamental result, the preceding analysis outlines an approach for characterizing the minimum dissipation to hold a system out of equilibrium.
We now carry out this analysis again in a new scenario, but allow the addition of $C$ extra nodes in the network and edges connecting them (Figure~\ref{fig:graph}).

When we add additional nodes, the system plus controller will have $\alpha=1,\dots, N+C$ configurations, with a steady-state distribution $\rho^{\rm ss}_\alpha$ over the super-system.
Now, in this case control will be successful when in the resulting steady state the relative likelihood of the $N$ original states are in the target distribution $p^*_i=\rho^{\rm ss}_i/{\mathcal P}$, where ${\mathcal P}=\sum_{i=1}^N \rho_i^{\rm ss}$.
Unfortunately, this is not sufficient to fix the dissipation rate on the original set of links, as the currents are left undetermined.
Indeed, it is possible to have the subset of $N$ system nodes in the target distribution $p^*$, but have ${\mathcal P}$ small; leading to small currents on the uncontrolled links $J_{ij}(\rho^{\rm ss})={\mathcal P}J_{ij}(p^*)$ and negligible dissipation [cf.~\eqref{eq:entProd}].
Thus to arrive at a sensible bound we must also fix  the probability currents $J_{ij}(\rho^{\rm ss})={\mathcal P}J_{ij}(p^*)$ on the original uncontrolled links, or equivalently ${\mathcal P}$, the total probability to be in the original configurations.
In effect, we are maintaining the function of the system, as the currents represent different possible tasks for the system, {\emph e.g.}, they are the rate of production of a molecule or the rate at which heat flux is converted into useful work.

With this setup, the minimum entropy production rate is again given by the entropy production on the original undriven links in the global steady state

\begin{equation}
{\dot S}_{\rm i}\ge {\dot S}_{\rm min} =k_{\rm B}\sum_{i>j} J_{ij}(\rho^{\rm ss})\ln \frac{W_{ij} \rho^{\rm ss}_{j}}{W_{ji}\rho^{\rm ss}_{i}}.
\end{equation}
To make this an expression that only depends on $p^*$ and ${\mathcal P}$, we substitute $\rho^{\rm ss}_i=p^*_i{\mathcal P}$ to find

\begin{align}
 {\dot S}_{\rm min} &=k_{\rm B}{\mathcal P}\sum_{i>j} J_{ij}(p^*)\ln \frac{W_{ij} p^*_{j}}{W_{ji}p^*_{i}}\\
&=-k_{\rm B}{\mathcal P}\partial_t^{\rm eq}D(p^*||p^{\rm eq}).
\end{align}
Again, the minimum dissipation is dictated by how different the target state is from equilibrium, but here weighted by the total probability ${\mathcal P}$, which fixes the system's currents.
Similarly, optimality is reached when the additional edges that connect the control nodes are very fast, minimizing their contribution to the entropy production.

\subsection{Auxiliary control}
Another scheme for control is the addition of an entirely new system, which we call the auxiliary.
This scheme was original analyzed in \cite{Horowitz2015b} for quantum mesoscopic devices modeled by a Markovian quantum Master equation.
In an effort to unify various results, we recapitulate this argument here, translated into classical language. 

We now amend our state space with the addition of an auxiliary control system with states $\alpha=1,\dots,C$, so that each configuration of the super-system is labeled by the pair $(i,\alpha)$.
Transition rates of the original system are assumed unaltered, but the new transitions between auxiliary states $\{M_{\alpha\gamma}^i\}$ must  depend on the system state in order to implement the feedback control.
Such a structure is called bipartite~\cite{Barato2013b, Diana2013b, Hartich2014, Horowitz2014}, and is captured in the graph structure by the absence of diagonal links where the system and auxiliary transition simultaneously (Figure~\ref{fig:graph}).
Here, control is successful when the steady-state distribution $\rho_{i\alpha}^{\rm ss}$ has a marginal distribution on the system that is the target distribution $\sum_{\alpha=1}^C\rho^{\rm ss}_{i\alpha}=p_i^*$.

Again we can bound the total entropy production in the system plus auxiliary with the entropy production on just the system links

\begin{equation}
{\dot S}_{\rm i}\ge k_{\rm B}\sum_{(i,\alpha)>(j,\gamma)}J_{ij}(\rho^{\rm ss})\ln \frac{W_{ij}\rho^{\rm ss}_{j \gamma}}{ W_{ji}\rho^{\rm ss}_{i\alpha}}.
\end{equation}
At this point the lower bound still depends on the full distribution $\rho^{\rm ss}$ over the entire super-system.
However, if we coarse-grain over the auxiliary, we can use the monotonicity of the relative entropy under coarse-graining~\cite{Cover}, to weaken the bound to

\begin{align}
{\dot S}_{\rm i}\ge {\dot S}_{\rm min}&= k_{\rm B}\sum_{i>j}J_{ij}(p^{*})\ln \frac{W_{ij}p^*_{j}}{ W_{ji}p^*_{i}}\\
&=-k_{\rm B}\partial_t^{\rm eq}D(p^*||p^{\rm eq}).
\end{align}
This result was originally derived in the context of quantum mesoscopoic devices~\cite{Horowitz2015b}.
Here we have reframed it in classical language.

\subsection{Controlling chemical reaction networks}
As a final scenario, we turn to the control of a chemical reaction network.
This scenario adds an additional complication: the incorporation of additional control reactions can break an underlying conservation law of the equilibrium dynamics~\cite{Polettini2014,Rao2016}. For example, adding a chemostat that exchanges matter breaks the conservation of particle number or mass.
This observation requires a slight modification of \eqref{eq:edgeBound2}.

\begin{figure}[tb]
\centering
\includegraphics[scale=.5]{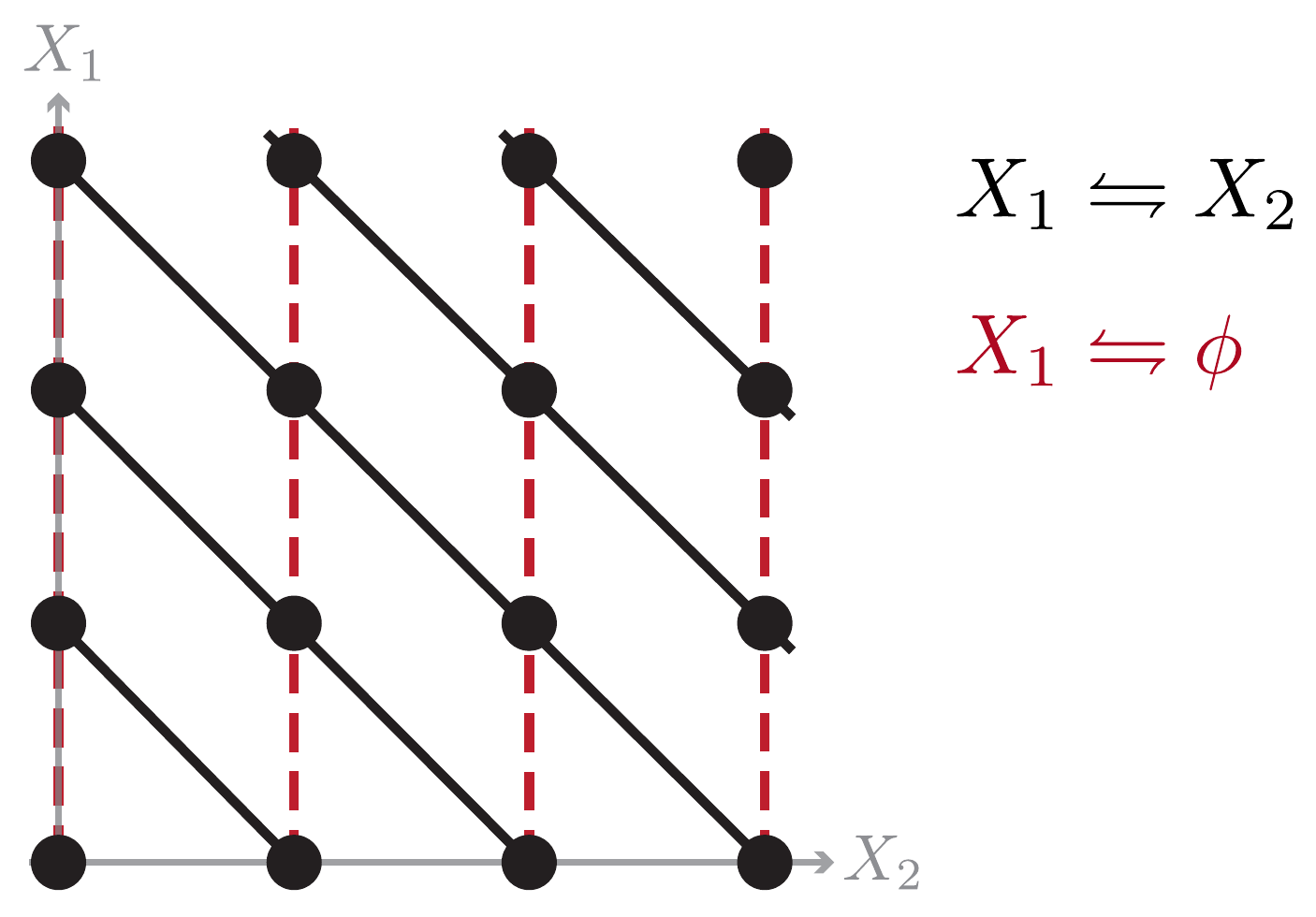}
\caption{Illustration of chemical control: Two species $X_1$ and $X_2$ interconvert through a single chemical reaction $X_1\rightleftharpoons X_2$ that conserves particle number, depicted as solid black lines.  As a result, the dynamical evolution of this chemical reaction network is restricted to a single diagonal subspace of the network of states.  Control can be implemented by chemostating one of the species, say $X_1$, by allowing $X_1$ molecules to be added and subtracted from the reaction volume through the reaction $X_1\rightleftharpoons \phi$, depicted as red dashed lines.  As this reaction breaks the particle number conservation law, it extends the possible configurations the system can dynamically explore.}
\label{fig:chem}
\end{figure}

To set the stage, consider a chemical reaction network with configurations specified by the vector of chemical species number ${\bf X}=\{X_1,\dots, X_S\}$.
Transitions then correspond to chemical reactions that change the value of ${\bf X}$ subject to system-specific constraints; an illustrative example of which is pictured in Figure~\ref{fig:chem}.
For simplicity, we take the only constraint to be particle number ${\mathcal N}=\sum_{i=1}^S X_i$.
In which case, the equilibrium steady state is a Poisson distribution constrained to the manifold of fixed particle number, $p_{\mathcal N}^{\rm eq}({\bf X})=\prod_i\frac{{\bar X}_i^{X_i}}{X_i!}e^{-{\bar X}_i}\delta(\sum_j X_j-{\mathcal N})$~\cite{VanKampen}; and detailed balance respects the constraints as well: 

\begin{equation}
W_{{\bf X}^\prime{\bf X}}p^{\rm eq}_{\mathcal N}({\bf X})=W_{{\bf X}{\bf X}^\prime}p^{\rm eq}_{\mathcal N}({\bf X}^\prime).
\end{equation}

Now, for control we add new reactions that maintain the system in a fixed target distribution $p^*({\bf X})$ over chemical space, which may not respect our particle number constraint, \emph{i.e.}, it may have support on configurations ${\bf X}$ that have different numbers of total particle number (cf. Figure~\ref{fig:chem}).
To make this explicit, we split the target distribution $p^*({\bf X})=p_{\mathcal N}^*({\bf X}){\mathcal P}_{\mathcal N}$ into two controlloable pieces: the conditional probability given the total particle number, $p_{\mathcal N}^*({\bf X})$ and the probability to have ${\mathcal N}$ particles, ${\mathcal P}_{\mathcal N}$.
With this splitting, the minimum dissipation to maintain $p^*$ again is only due to the entropy produced in the original reactions

\begin{align}
{\dot S}_{\rm min}&=k_{\rm B} \sum_{{\bf X}^\prime>{\bf X}} J_{{\bf X}^\prime{\bf X}}(p^*)\ln\frac{W_{{\bf X}^\prime{\bf X}}p^*({\bf X})}{W_{{\bf X}{\bf X}^\prime}p^*({\bf X}^\prime)} \\
&=k_{\rm B}\sum_{\mathcal N}{\mathcal P}_{\mathcal N} \sum_{{\bf X}^\prime>{\bf X}} J_{{\bf X}^\prime{\bf X}}(p^*_{\mathcal N})\ln\frac{W_{{\bf X}^\prime{\bf X}}p_{\mathcal N}^*({\bf X})}{W_{{\bf X}{\bf X}^\prime}p_{\mathcal N}^*({\bf X}^\prime)},
\end{align}
An information-theoretic interpretation is provided by recalling that the equilibrium transitions $W_{{\bf X}^\prime{\bf X}}$ conserve particle number, to find

\begin{equation}
 {\dot S}_{\rm min}= -k_{\rm B}\sum_{\mathcal N}{\mathcal P}_{\mathcal N}\partial^{\rm eq}_tD(p^*_{\mathcal N}||p_{\mathcal N}^{\rm eq}).
\end{equation}
In a chemical system, the minimum dissipation depends only on how different the target distribution is from the equilibrium distribution on the conserved sectors, whereas shifting only the number distribution ${\mathcal P}_{\mathcal N}$ can in principle be accomplished for free.
This conclusion should remain true when there are additional conservation laws as well.
Note that when the target distribution conserves particle number, ${\mathcal P}_{\mathcal N}=1$ and we recover \eqref{eq:edgeBound2}.

\section{Discussion}

We investigated the minimum entropy production or free energy dissipation to maintain a system in a target nonequilibrium distribution using ancillary control.
We found that in a variety of scenarios this minimum cost can be formulated using the information-theoretic relative entropy as a measure of how distinguishable the target nonequilibrium state is from equilibrium.
Our analysis further revealed that the minimum is reached when the driven control transitions operated reversibly.

As in previous analyses of nonequilibrium thermodynamics the relative entropy   \cite{Kawai2007, Parrondo2009, Spohn1978, Ge2010, Gaveau1997,Procaccia1976} appeared as a key tool in characterizing dissipation.
In these previous works, however, the relative entropy compared the true evolution of the system to the underlying stationary state.
By contrast, here we find that when using external control the cost is characterized by the time-variation of the relative entropy under a fictitious uncontrolled equilibrium dynamics, evaluated against the unperturbed equilibrium state.

Looking ahead, we note that while we had in mind throughout the paper autonomous control, nonautonomous control through reversible hidden pumps offer an intriguing alternative to saturate our energetic bound~\cite{Esposito2014}.
Additionally, we have focused on the average or typical behavior of the state of our system, but a number recent predictions, collectively known as thermodynamic uncertainty relations, relate the dissipation to fluctuations in currents or flows~\cite{Barato2015,Pietzonka2016Affinity,Pietzonka2016Universal,Gingrich2016,Gingrich2017}.
Such work suggests that it would be intriguing to understand how our lower bound is modified, when one wants to use external control to constrain not just the typical state, but fluctuations as well.
Our hope is that the approach developed here offers the possibility of quantifying the minimum energetic cost of nonequilibrium states in other more general scenarios.

\acknowledgments{This work was supported by the Gordon and Betty Moore Foundation through Grant GBMF4343. J.L.E. further acknowledges the Cabot family for their generous support of MIT.}

\bibliography{FluctuationTheory.bib,PhysicsTexts.bib,Feedback.bib}

\begin{thebibliography}{45}%
\makeatletter
\providecommand \@ifxundefined [1]{%
 \@ifx{#1\undefined}
}%
\providecommand \@ifnum [1]{%
 \ifnum #1\expandafter \@firstoftwo
 \else \expandafter \@secondoftwo
 \fi
}%
\providecommand \@ifx [1]{%
 \ifx #1\expandafter \@firstoftwo
 \else \expandafter \@secondoftwo
 \fi
}%
\providecommand \natexlab [1]{#1}%
\providecommand \enquote  [1]{``#1''}%
\providecommand \bibnamefont  [1]{#1}%
\providecommand \bibfnamefont [1]{#1}%
\providecommand \citenamefont [1]{#1}%
\providecommand \href@noop [0]{\@secondoftwo}%
\providecommand \href [0]{\begingroup \@sanitize@url \@href}%
\providecommand \@href[1]{\@@startlink{#1}\@@href}%
\providecommand \@@href[1]{\endgroup#1\@@endlink}%
\providecommand \@sanitize@url [0]{\catcode `\\12\catcode `\$12\catcode
  `\&12\catcode `\#12\catcode `\^12\catcode `\_12\catcode `\%12\relax}%
\providecommand \@@startlink[1]{}%
\providecommand \@@endlink[0]{}%
\providecommand \url  [0]{\begingroup\@sanitize@url \@url }%
\providecommand \@url [1]{\endgroup\@href {#1}{\urlprefix }}%
\providecommand \urlprefix  [0]{URL }%
\providecommand \Eprint [0]{\href }%
\@ifxundefined \urlstyle {%
  \providecommand \doi  [0]{\begingroup \@sanitize@url \@doi}%
  \providecommand \@doi [1]{\endgroup \@@startlink {\doibase
  #1}doi:\discretionary {}{}{}#1\@@endlink }%
}{%
  \providecommand \doi  [0]{doi:\discretionary{}{}{}\begingroup
  \urlstyle{rm}\Url }%
}%
\providecommand \doibase [0]{http://dx.doi.org/}%
\providecommand \Doi [0]{\begingroup \@sanitize@url \@Doi }%
\providecommand \@Doi  [1]{\endgroup\@@startlink{\doibase#1}\@@Doi}%
\providecommand \@@Doi [1]{#1\@@endlink}%
\providecommand \selectlanguage [0]{\@gobble}%
\providecommand \bibinfo  [0]{\@secondoftwo}%
\providecommand \bibfield  [0]{\@secondoftwo}%
\providecommand \translation [1]{[#1]}%
\providecommand \BibitemOpen [0]{}%
\providecommand \bibitemStop [0]{}%
\providecommand \bibitemNoStop [0]{.\EOS\space}%
\providecommand \EOS [0]{\spacefactor3000\relax}%
\providecommand \BibitemShut  [1]{\csname bibitem#1\endcsname}%
\bibitem [{\citenamefont {Parrondo}\ and\ \citenamefont
  {De~Cisneros}(2002)}]{Parrondo2002}%
  \BibitemOpen
  \bibfield  {author} {\bibinfo {author} {\bibfnamefont {J.~M.~R.}\
  \bibnamefont {Parrondo}}\ and\ \bibinfo {author} {\bibfnamefont {B.~J.}\
  \bibnamefont {De~Cisneros}},\ }\href@noop {} {\bibfield  {journal} {\bibinfo
  {journal} {Appl. Phys. A},\ }\textbf {\bibinfo {volume} {75}},\ \bibinfo
  {pages} {179} (\bibinfo {year} {2002})}\BibitemShut {NoStop}%
\bibitem [{\citenamefont {Bialek}(2012)}]{Bialek}%
  \BibitemOpen
  \bibfield  {author} {\bibinfo {author} {\bibfnamefont {W.}~\bibnamefont
  {Bialek}},\ }\href@noop {} {\emph {\bibinfo {title} {Biophysics: Searching
  for principles}}}\ (\bibinfo  {publisher} {Princeton University Press, New
  Jersey},\ \bibinfo {year} {2012})\BibitemShut {NoStop}%
\bibitem [{\citenamefont {Li}\ \emph {et~al.}(2011)\citenamefont {Li},
  \citenamefont {Kheifets},\ and\ \citenamefont {Raizen}}]{Li2011}%
  \BibitemOpen
  \bibfield  {author} {\bibinfo {author} {\bibfnamefont {T.}~\bibnamefont
  {Li}}, \bibinfo {author} {\bibfnamefont {S.}~\bibnamefont {Kheifets}}, \ and\
  \bibinfo {author} {\bibfnamefont {M.}~\bibnamefont {Raizen}},\ }\href@noop {}
  {\bibfield  {journal} {\bibinfo  {journal} {Nat. Phys.},\ }\textbf {\bibinfo
  {volume} {7}},\ \bibinfo {pages} {527} (\bibinfo {year} {2011})}\BibitemShut
  {NoStop}%
\bibitem [{\citenamefont {Tian}(2009)}]{Tian09}%
  \BibitemOpen
  \bibfield  {author} {\bibinfo {author} {\bibfnamefont {L.}~\bibnamefont
  {Tian}},\ }\Doi {10.1103/PhysRevB.79.193407} {\bibfield  {journal} {\bibinfo
  {journal} {Phys. Rev. B},\ }\textbf {\bibinfo {volume} {79}},\ \bibinfo
  {pages} {193407} (\bibinfo {year} {2009})}\BibitemShut {NoStop}%
\bibitem [{\citenamefont {Palomaki}\ \emph {et~al.}(2013)\citenamefont
  {Palomaki}, \citenamefont {Harlow}, \citenamefont {Teufel}, \citenamefont
  {Simmonds},\ and\ \citenamefont {Lehnert}}]{Palomaki13}%
  \BibitemOpen
  \bibfield  {author} {\bibinfo {author} {\bibfnamefont {T.~A.}\ \bibnamefont
  {Palomaki}}, \bibinfo {author} {\bibfnamefont {J.~W.}\ \bibnamefont
  {Harlow}}, \bibinfo {author} {\bibfnamefont {J.~D.}\ \bibnamefont {Teufel}},
  \bibinfo {author} {\bibfnamefont {R.~W.}\ \bibnamefont {Simmonds}}, \ and\
  \bibinfo {author} {\bibfnamefont {K.~W.}\ \bibnamefont {Lehnert}},\ }\Doi
  {doi:10.1038/nature11915} {\bibfield  {journal} {\bibinfo  {journal}
  {Nature},\ }\textbf {\bibinfo {volume} {495}},\ \bibinfo {pages} {210}
  (\bibinfo {year} {2013})}\BibitemShut {NoStop}%
\bibitem [{\citenamefont {Horowitz}\ \emph {et~al.}(2013)\citenamefont
  {Horowitz}, \citenamefont {Sagawa},\ and\ \citenamefont
  {Parrondo}}]{Horowitz2013}%
  \BibitemOpen
  \bibfield  {author} {\bibinfo {author} {\bibfnamefont {J.~M.}\ \bibnamefont
  {Horowitz}}, \bibinfo {author} {\bibfnamefont {T.}~\bibnamefont {Sagawa}}, \
  and\ \bibinfo {author} {\bibfnamefont {J.~M.~R.}\ \bibnamefont {Parrondo}},\
  }\href@noop {} {\bibfield  {journal} {\bibinfo  {journal} {Phys. Rev.
  Lett.},\ }\textbf {\bibinfo {volume} {111}},\ \bibinfo {pages} {010602}
  (\bibinfo {year} {2013})}\BibitemShut {NoStop}%
\bibitem [{\citenamefont {Munakata}\ and\ \citenamefont
  {Rosinberg}(2012)}]{Munakata2012}%
  \BibitemOpen
  \bibfield  {author} {\bibinfo {author} {\bibfnamefont {T.}~\bibnamefont
  {Munakata}}\ and\ \bibinfo {author} {\bibfnamefont {M.~L.}\ \bibnamefont
  {Rosinberg}},\ }\href@noop {} {\bibfield  {journal} {\bibinfo  {journal} {J.
  Stat. Mech.},\ \bibinfo {pages} {P05010}} (\bibinfo {year}
  {2012})}\BibitemShut {NoStop}%
\bibitem [{\citenamefont {Parrondo}\ \emph {et~al.}(2015)\citenamefont
  {Parrondo}, \citenamefont {Horowitz},\ and\ \citenamefont
  {Sagawa}}]{Parrondo2015}%
  \BibitemOpen
  \bibfield  {author} {\bibinfo {author} {\bibfnamefont {J.~M.~R.}\
  \bibnamefont {Parrondo}}, \bibinfo {author} {\bibfnamefont {J.~M.}\
  \bibnamefont {Horowitz}}, \ and\ \bibinfo {author} {\bibfnamefont
  {T.}~\bibnamefont {Sagawa}},\ }\href@noop {} {\bibfield  {journal} {\bibinfo
  {journal} {Nature Phys.},\ }\textbf {\bibinfo {volume} {11}},\ \bibinfo
  {pages} {131} (\bibinfo {year} {2015})}\BibitemShut {NoStop}%
\bibitem [{\citenamefont {Sandberg}\ \emph {et~al.}(2014)\citenamefont
  {Sandberg}, \citenamefont {Delvenne}, \citenamefont {Newton},\ and\
  \citenamefont {Mitter}}]{Sandberg2014}%
  \BibitemOpen
  \bibfield  {author} {\bibinfo {author} {\bibfnamefont {H.}~\bibnamefont
  {Sandberg}}, \bibinfo {author} {\bibfnamefont {J.-C.}\ \bibnamefont
  {Delvenne}}, \bibinfo {author} {\bibfnamefont {N.~J.}\ \bibnamefont
  {Newton}}, \ and\ \bibinfo {author} {\bibfnamefont {S.~K.}\ \bibnamefont
  {Mitter}},\ }\href@noop {} {\bibfield  {journal} {\bibinfo  {journal} {Phys.
  Rev. E},\ }\textbf {\bibinfo {volume} {90}},\ \bibinfo {pages} {042119}
  (\bibinfo {year} {2014})}\BibitemShut {NoStop}%
\bibitem [{\citenamefont {Horowitz}\ and\ \citenamefont
  {Esposito}(2014)}]{Horowitz2014}%
  \BibitemOpen
  \bibfield  {author} {\bibinfo {author} {\bibfnamefont {J.~M.}\ \bibnamefont
  {Horowitz}}\ and\ \bibinfo {author} {\bibfnamefont {M.}~\bibnamefont
  {Esposito}},\ }\href@noop {} {\bibfield  {journal} {\bibinfo  {journal}
  {Phys. Rev. X},\ }\textbf {\bibinfo {volume} {4}},\ \bibinfo {pages} {031015}
  (\bibinfo {year} {2014})}\BibitemShut {NoStop}%
\bibitem [{\citenamefont {Horowitz}\ and\ \citenamefont
  {Sandberg}(2014)}]{Horowitz2014b}%
  \BibitemOpen
  \bibfield  {author} {\bibinfo {author} {\bibfnamefont {J.~M.}\ \bibnamefont
  {Horowitz}}\ and\ \bibinfo {author} {\bibfnamefont {H.}~\bibnamefont
  {Sandberg}},\ }\href@noop {} {\bibfield  {journal} {\bibinfo  {journal} {New
  J. Phys.},\ }\textbf {\bibinfo {volume} {15}},\ \bibinfo {pages} {125007}
  (\bibinfo {year} {2014})}\BibitemShut {NoStop}%
\bibitem [{\citenamefont {Shiraishi}\ and\ \citenamefont
  {Sagawa}(2015)}]{Shiraishi2014}%
  \BibitemOpen
  \bibfield  {author} {\bibinfo {author} {\bibfnamefont {N.}~\bibnamefont
  {Shiraishi}}\ and\ \bibinfo {author} {\bibfnamefont {T.}~\bibnamefont
  {Sagawa}},\ }\href@noop {} {\bibfield  {journal} {\bibinfo  {journal} {Phys.
  Rev. E},\ }\textbf {\bibinfo {volume} {91}},\ \bibinfo {pages} {012130}
  (\bibinfo {year} {2015})}\BibitemShut {NoStop}%
\bibitem [{\citenamefont {Kondepudi}\ and\ \citenamefont
  {Prigogine}(2014)}]{Kondepudi}%
  \BibitemOpen
  \bibfield  {author} {\bibinfo {author} {\bibfnamefont {D.}~\bibnamefont
  {Kondepudi}}\ and\ \bibinfo {author} {\bibfnamefont {I.}~\bibnamefont
  {Prigogine}},\ }\href@noop {} {\emph {\bibinfo {title} {Modern
  Thermodynamics: From heat engines to dissipative structures}}},\ \bibinfo
  {edition} {2nd}\ ed.\ (\bibinfo  {publisher} {John Wiley \& Sons, Ltd,
  Chichester, UK},\ \bibinfo {year} {2014})\BibitemShut {NoStop}%
\bibitem [{\citenamefont {Maes}\ and\ \citenamefont
  {Neto\v{c}n\'{y}}(2007)}]{Maes2007}%
  \BibitemOpen
  \bibfield  {author} {\bibinfo {author} {\bibfnamefont {C.}~\bibnamefont
  {Maes}}\ and\ \bibinfo {author} {\bibfnamefont {K.}~\bibnamefont
  {Neto\v{c}n\'{y}}},\ }\href@noop {} {\bibfield  {journal} {\bibinfo
  {journal} {J. Math. Phys.},\ }\textbf {\bibinfo {volume} {48}},\ \bibinfo
  {pages} {053306} (\bibinfo {year} {2007})}\BibitemShut {NoStop}%
\bibitem [{\citenamefont {Bruers}\ \emph {et~al.}(2007)\citenamefont {Bruers},
  \citenamefont {Maes},\ and\ \citenamefont {Neto\v{c}n\'{y}}}]{Bruers2007}%
  \BibitemOpen
  \bibfield  {author} {\bibinfo {author} {\bibfnamefont {S.}~\bibnamefont
  {Bruers}}, \bibinfo {author} {\bibfnamefont {C.}~\bibnamefont {Maes}}, \ and\
  \bibinfo {author} {\bibfnamefont {K.}~\bibnamefont {Neto\v{c}n\'{y}}},\
  }\href@noop {} {\bibfield  {journal} {\bibinfo  {journal} {J. Stat. Phys.},\
  }\textbf {\bibinfo {volume} {129}},\ \bibinfo {pages} {725} (\bibinfo {year}
  {2007})}\BibitemShut {NoStop}%
\bibitem [{\citenamefont {Polettini}\ and\ \citenamefont
  {Esposito}(2013)}]{Poletinni2013}%
  \BibitemOpen
  \bibfield  {author} {\bibinfo {author} {\bibfnamefont {M.}~\bibnamefont
  {Polettini}}\ and\ \bibinfo {author} {\bibfnamefont {M.}~\bibnamefont
  {Esposito}},\ }\href@noop {} {\bibfield  {journal} {\bibinfo  {journal}
  {Phys. Rev. E},\ }\textbf {\bibinfo {volume} {88}},\ \bibinfo {pages}
  {012112} (\bibinfo {year} {2013})}\BibitemShut {NoStop}%
\bibitem [{\citenamefont {Horowitz}\ \emph {et~al.}(2017)\citenamefont
  {Horowitz}, \citenamefont {Zhou},\ and\ \citenamefont
  {England}}]{Horowitz2017}%
  \BibitemOpen
  \bibfield  {author} {\bibinfo {author} {\bibfnamefont {J.~M.}\ \bibnamefont
  {Horowitz}}, \bibinfo {author} {\bibfnamefont {K.}~\bibnamefont {Zhou}}, \
  and\ \bibinfo {author} {\bibfnamefont {J.~L.}\ \bibnamefont {England}},\
  }\href@noop {} {\bibfield  {journal} {\bibinfo  {journal} {Phys. Rev. E},\
  }\textbf {\bibinfo {volume} {95}},\ \bibinfo {pages} {042102} (\bibinfo
  {year} {2017})}\BibitemShut {NoStop}%
\bibitem [{\citenamefont {Horowitz}\ and\ \citenamefont
  {Jacobs}(2015)}]{Horowitz2015b}%
  \BibitemOpen
  \bibfield  {author} {\bibinfo {author} {\bibfnamefont {J.~M.}\ \bibnamefont
  {Horowitz}}\ and\ \bibinfo {author} {\bibfnamefont {K.}~\bibnamefont
  {Jacobs}},\ }\href@noop {} {\bibfield  {journal} {\bibinfo  {journal} {Phys.
  Rev. Lett.},\ }\textbf {\bibinfo {volume} {115}},\ \bibinfo {pages} {130501}
  (\bibinfo {year} {2015})}\BibitemShut {NoStop}%
\bibitem [{\citenamefont {Van~Kampen}(2007)}]{VanKampen}%
  \BibitemOpen
  \bibfield  {author} {\bibinfo {author} {\bibfnamefont {N.~G.}\ \bibnamefont
  {Van~Kampen}},\ }\href@noop {} {\emph {\bibinfo {title} {Stochastic Processes
  in Physics and Chemistry}}},\ \bibinfo {edition} {3rd}\ ed.\ (\bibinfo
  {publisher} {Elsevier Ltd., New York},\ \bibinfo {year} {2007})\BibitemShut
  {NoStop}%
\bibitem [{\citenamefont {Esposito}\ and\ \citenamefont {Van~den
  Broeck}(2011)}]{Esposito2011}%
  \BibitemOpen
  \bibfield  {author} {\bibinfo {author} {\bibfnamefont {M.}~\bibnamefont
  {Esposito}}\ and\ \bibinfo {author} {\bibfnamefont {C.}~\bibnamefont {Van~den
  Broeck}},\ }\href@noop {} {\bibfield  {journal} {\bibinfo  {journal}
  {Europhys. Lett.},\ }\textbf {\bibinfo {volume} {95}},\ \bibinfo {pages}
  {40004} (\bibinfo {year} {2011})}\BibitemShut {NoStop}%
\bibitem [{\citenamefont {Sartori}\ \emph {et~al.}(2014)\citenamefont
  {Sartori}, \citenamefont {Granger}, \citenamefont {Lee},\ and\ \citenamefont
  {Horowitz}}]{Sartori2014}%
  \BibitemOpen
  \bibfield  {author} {\bibinfo {author} {\bibfnamefont {P.}~\bibnamefont
  {Sartori}}, \bibinfo {author} {\bibfnamefont {L.}~\bibnamefont {Granger}},
  \bibinfo {author} {\bibfnamefont {C.}~\bibnamefont {Lee}}, \ and\ \bibinfo
  {author} {\bibfnamefont {J.}~\bibnamefont {Horowitz}},\ }\href@noop {}
  {\bibfield  {journal} {\bibinfo  {journal} {PloS Comput. Biol.},\ }\textbf
  {\bibinfo {volume} {10}},\ \bibinfo {pages} {e1003974} (\bibinfo {year}
  {2014})}\BibitemShut {NoStop}%
\bibitem [{\citenamefont {Sartori}\ and\ \citenamefont
  {Piglotti}(2015)}]{Sartori2015}%
  \BibitemOpen
  \bibfield  {author} {\bibinfo {author} {\bibfnamefont {P.}~\bibnamefont
  {Sartori}}\ and\ \bibinfo {author} {\bibfnamefont {S.}~\bibnamefont
  {Piglotti}},\ }\href@noop {} {\bibfield  {journal} {\bibinfo  {journal}
  {Phys. Rev. X},\ }\textbf {\bibinfo {volume} {5}},\ \bibinfo {pages} {041039}
  (\bibinfo {year} {2015})}\BibitemShut {NoStop}%
\bibitem [{\citenamefont {Horowitz}\ and\ \citenamefont
  {Jacobs}(2014)}]{Horowitz2013c}%
  \BibitemOpen
  \bibfield  {author} {\bibinfo {author} {\bibfnamefont {J.~M.}\ \bibnamefont
  {Horowitz}}\ and\ \bibinfo {author} {\bibfnamefont {K.}~\bibnamefont
  {Jacobs}},\ }\href@noop {} {\bibfield  {journal} {\bibinfo  {journal} {Phys.
  Rev. E},\ }\textbf {\bibinfo {volume} {89}},\ \bibinfo {pages} {042134}
  (\bibinfo {year} {2014})}\BibitemShut {NoStop}%
\bibitem [{\citenamefont {Hamerly}\ and\ \citenamefont
  {Mabuchi}(2012)}]{Hamerly2012}%
  \BibitemOpen
  \bibfield  {author} {\bibinfo {author} {\bibfnamefont {R.}~\bibnamefont
  {Hamerly}}\ and\ \bibinfo {author} {\bibfnamefont {H.}~\bibnamefont
  {Mabuchi}},\ }\href@noop {} {\bibfield  {journal} {\bibinfo  {journal} {Phys.
  Rev. Lett.},\ }\textbf {\bibinfo {volume} {109}},\ \bibinfo {pages} {173602}
  (\bibinfo {year} {2012})}\BibitemShut {NoStop}%
\bibitem [{\citenamefont {Schliesser}\ \emph {et~al.}(2008)\citenamefont
  {Schliesser}, \citenamefont {Rivi\`{e}re}, \citenamefont {Anetsberger},
  \citenamefont {Arcizet},\ and\ \citenamefont {Kippenberg}}]{Schliesser08}%
  \BibitemOpen
  \bibfield  {author} {\bibinfo {author} {\bibfnamefont {A.}~\bibnamefont
  {Schliesser}}, \bibinfo {author} {\bibfnamefont {R.}~\bibnamefont
  {Rivi\`{e}re}}, \bibinfo {author} {\bibfnamefont {G.}~\bibnamefont
  {Anetsberger}}, \bibinfo {author} {\bibfnamefont {O.}~\bibnamefont
  {Arcizet}}, \ and\ \bibinfo {author} {\bibfnamefont {T.~J.}\ \bibnamefont
  {Kippenberg}},\ }\Doi {10.1038/nphys939} {\bibfield  {journal} {\bibinfo
  {journal} {Nature Phys.},\ }\textbf {\bibinfo {volume} {5}},\ \bibinfo
  {pages} {415} (\bibinfo {year} {2008})}\BibitemShut {NoStop}%
\bibitem [{\citenamefont {Esposito}\ and\ \citenamefont {Van~den
  Broeck}(2010)}]{Esposito2010b}%
  \BibitemOpen
  \bibfield  {author} {\bibinfo {author} {\bibfnamefont {M.}~\bibnamefont
  {Esposito}}\ and\ \bibinfo {author} {\bibfnamefont {C.}~\bibnamefont {Van~den
  Broeck}},\ }\href@noop {} {\bibfield  {journal} {\bibinfo  {journal} {Phys.
  Rev. E},\ }\textbf {\bibinfo {volume} {82}},\ \bibinfo {pages} {011143}
  (\bibinfo {year} {2010})}\BibitemShut {NoStop}%
\bibitem [{\citenamefont {Seifert}(2012)}]{Seifert2012}%
  \BibitemOpen
  \bibfield  {author} {\bibinfo {author} {\bibfnamefont {U.}~\bibnamefont
  {Seifert}},\ }\href@noop {} {\bibfield  {journal} {\bibinfo  {journal} {Rep.
  Prog. Phys.},\ }\textbf {\bibinfo {volume} {75}},\ \bibinfo {pages} {126001}
  (\bibinfo {year} {2012})}\BibitemShut {NoStop}%
\bibitem [{\citenamefont {Cover}\ and\ \citenamefont {Thomas}(2006)}]{Cover}%
  \BibitemOpen
  \bibfield  {author} {\bibinfo {author} {\bibfnamefont {T.~M.}\ \bibnamefont
  {Cover}}\ and\ \bibinfo {author} {\bibfnamefont {J.~A.}\ \bibnamefont
  {Thomas}},\ }\href@noop {} {\emph {\bibinfo {title} {Elements of Information
  Theory}}},\ \bibinfo {edition} {2nd}\ ed.\ (\bibinfo  {publisher}
  {Wiley-Interscience, New York},\ \bibinfo {year} {2006})\BibitemShut
  {NoStop}%
\bibitem [{\citenamefont {Barato}\ \emph {et~al.}(2013)\citenamefont {Barato},
  \citenamefont {Hartich},\ and\ \citenamefont {Seifert}}]{Barato2013b}%
  \BibitemOpen
  \bibfield  {author} {\bibinfo {author} {\bibfnamefont {A.~C.}\ \bibnamefont
  {Barato}}, \bibinfo {author} {\bibfnamefont {D.}~\bibnamefont {Hartich}}, \
  and\ \bibinfo {author} {\bibfnamefont {U.}~\bibnamefont {Seifert}},\
  }\href@noop {} {\bibfield  {journal} {\bibinfo  {journal} {J. Stat. Phys.},\
  }\textbf {\bibinfo {volume} {153}},\ \bibinfo {pages} {460} (\bibinfo {year}
  {2013})}\BibitemShut {NoStop}%
\bibitem [{\citenamefont {Diana}\ and\ \citenamefont
  {Esposito}(2014)}]{Diana2013b}%
  \BibitemOpen
  \bibfield  {author} {\bibinfo {author} {\bibfnamefont {G.}~\bibnamefont
  {Diana}}\ and\ \bibinfo {author} {\bibfnamefont {M.}~\bibnamefont
  {Esposito}},\ }\href@noop {} {\bibfield  {journal} {\bibinfo  {journal} {J.
  Stat. Mech.: Theor. Exp.},\ \bibinfo {pages} {P04010}} (\bibinfo {year}
  {2014})}\BibitemShut {NoStop}%
\bibitem [{\citenamefont {Hartich}\ \emph {et~al.}(2014)\citenamefont
  {Hartich}, \citenamefont {Barato},\ and\ \citenamefont
  {Seifert}}]{Hartich2014}%
  \BibitemOpen
  \bibfield  {author} {\bibinfo {author} {\bibfnamefont {D.}~\bibnamefont
  {Hartich}}, \bibinfo {author} {\bibfnamefont {A.~C.}\ \bibnamefont {Barato}},
  \ and\ \bibinfo {author} {\bibfnamefont {U.}~\bibnamefont {Seifert}},\
  }\href@noop {} {\bibfield  {journal} {\bibinfo  {journal} {J. Stat. Mech.},\
  \bibinfo {pages} {P02016}} (\bibinfo {year} {2014})}\BibitemShut {NoStop}%
\bibitem [{\citenamefont {Polettini}\ and\ \citenamefont
  {Esposito}(2014)}]{Polettini2014}%
  \BibitemOpen
  \bibfield  {author} {\bibinfo {author} {\bibfnamefont {M.}~\bibnamefont
  {Polettini}}\ and\ \bibinfo {author} {\bibfnamefont {M.}~\bibnamefont
  {Esposito}},\ }\href@noop {} {\bibfield  {journal} {\bibinfo  {journal} {J.
  Chem. Phys.},\ }\textbf {\bibinfo {volume} {141}},\ \bibinfo {pages} {024117}
  (\bibinfo {year} {2014})}\BibitemShut {NoStop}%
\bibitem [{\citenamefont {Rao}\ and\ \citenamefont {Esposito}(2016)}]{Rao2016}%
  \BibitemOpen
  \bibfield  {author} {\bibinfo {author} {\bibfnamefont {R.}~\bibnamefont
  {Rao}}\ and\ \bibinfo {author} {\bibfnamefont {M.}~\bibnamefont {Esposito}},\
  }\href@noop {} {\bibfield  {journal} {\bibinfo  {journal} {Phys. Rev. X},\
  }\textbf {\bibinfo {volume} {6}},\ \bibinfo {pages} {041064} (\bibinfo {year}
  {2016})}\BibitemShut {NoStop}%
\bibitem [{\citenamefont {Kawai}\ \emph {et~al.}(2007)\citenamefont {Kawai},
  \citenamefont {Parrondo},\ and\ \citenamefont {Van~den Broeck}}]{Kawai2007}%
  \BibitemOpen
  \bibfield  {author} {\bibinfo {author} {\bibfnamefont {R.}~\bibnamefont
  {Kawai}}, \bibinfo {author} {\bibfnamefont {J.~M.~R.}\ \bibnamefont
  {Parrondo}}, \ and\ \bibinfo {author} {\bibfnamefont {C.}~\bibnamefont
  {Van~den Broeck}},\ }\href@noop {} {\bibfield  {journal} {\bibinfo  {journal}
  {Phys. Rev. Lett.},\ }\textbf {\bibinfo {volume} {98}},\ \bibinfo {pages}
  {080602} (\bibinfo {year} {2007})}\BibitemShut {NoStop}%
\bibitem [{\citenamefont {Parrondo}\ \emph {et~al.}(2009)\citenamefont
  {Parrondo}, \citenamefont {Van~den Broeck},\ and\ \citenamefont
  {Kawai}}]{Parrondo2009}%
  \BibitemOpen
  \bibfield  {author} {\bibinfo {author} {\bibfnamefont {J.~M.~R.}\
  \bibnamefont {Parrondo}}, \bibinfo {author} {\bibfnamefont {C.}~\bibnamefont
  {Van~den Broeck}}, \ and\ \bibinfo {author} {\bibfnamefont {R.}~\bibnamefont
  {Kawai}},\ }\href@noop {} {\bibfield  {journal} {\bibinfo  {journal} {New J.
  Phys.},\ }\textbf {\bibinfo {volume} {11}},\ \bibinfo {pages} {073008}
  (\bibinfo {year} {2009})}\BibitemShut {NoStop}%
\bibitem [{\citenamefont {Spohn}\ and\ \citenamefont
  {Lebowitz}(1978)}]{Spohn1978}%
  \BibitemOpen
  \bibfield  {author} {\bibinfo {author} {\bibfnamefont {H.}~\bibnamefont
  {Spohn}}\ and\ \bibinfo {author} {\bibfnamefont {J.~L.}\ \bibnamefont
  {Lebowitz}},\ }in\ \href@noop {} {\emph {\bibinfo {booktitle} {Advances in
  Chemical Physics: For Ilya Prigogine}}},\ Vol.~\bibinfo {volume} {38},\
  \bibinfo {editor} {edited by\ \bibinfo {editor} {\bibfnamefont {S.~A.}\
  \bibnamefont {Rice}}}\ (\bibinfo  {publisher} {John Wiley \& Sons, Hoboken,
  NJ},\ \bibinfo {year} {1978})\BibitemShut {NoStop}%
\bibitem [{\citenamefont {Ge}\ and\ \citenamefont {Qian}(2010)}]{Ge2010}%
  \BibitemOpen
  \bibfield  {author} {\bibinfo {author} {\bibfnamefont {H.}~\bibnamefont
  {Ge}}\ and\ \bibinfo {author} {\bibfnamefont {H.}~\bibnamefont {Qian}},\
  }\href@noop {} {\bibfield  {journal} {\bibinfo  {journal} {Phys. Rev. E},\
  }\textbf {\bibinfo {volume} {81}},\ \bibinfo {pages} {051133} (\bibinfo
  {year} {2010})}\BibitemShut {NoStop}%
\bibitem [{\citenamefont {Gaveau}\ and\ \citenamefont
  {Schulman}(1997)}]{Gaveau1997}%
  \BibitemOpen
  \bibfield  {author} {\bibinfo {author} {\bibfnamefont {B.}~\bibnamefont
  {Gaveau}}\ and\ \bibinfo {author} {\bibfnamefont {L.~S.}\ \bibnamefont
  {Schulman}},\ }\href@noop {} {\bibfield  {journal} {\bibinfo  {journal}
  {Phys. Lett. A},\ }\textbf {\bibinfo {volume} {229}},\ \bibinfo {pages} {347}
  (\bibinfo {year} {1997})}\BibitemShut {NoStop}%
\bibitem [{\citenamefont {Procaccia}\ and\ \citenamefont
  {Levine}(1976)}]{Procaccia1976}%
  \BibitemOpen
  \bibfield  {author} {\bibinfo {author} {\bibfnamefont {I.}~\bibnamefont
  {Procaccia}}\ and\ \bibinfo {author} {\bibfnamefont {R.~D.}\ \bibnamefont
  {Levine}},\ }\href@noop {} {\bibfield  {journal} {\bibinfo  {journal} {J.
  Chem. Phys.},\ }\textbf {\bibinfo {volume} {65}},\ \bibinfo {pages} {3357}
  (\bibinfo {year} {1976})}\BibitemShut {NoStop}%
\bibitem [{\citenamefont {Esposito}\ and\ \citenamefont
  {Parrondo}(2015)}]{Esposito2014}%
  \BibitemOpen
  \bibfield  {author} {\bibinfo {author} {\bibfnamefont {M.}~\bibnamefont
  {Esposito}}\ and\ \bibinfo {author} {\bibfnamefont {J.~M.~R.}\ \bibnamefont
  {Parrondo}},\ }\href@noop {} {\bibfield  {journal} {\bibinfo  {journal}
  {Phys. Rev. E},\ }\textbf {\bibinfo {volume} {91}},\ \bibinfo {pages}
  {052114} (\bibinfo {year} {2015})}\BibitemShut {NoStop}%
\bibitem [{\citenamefont {Barato}\ and\ \citenamefont
  {Seifert}(2015)}]{Barato2015}%
  \BibitemOpen
  \bibfield  {author} {\bibinfo {author} {\bibfnamefont {A.~C.}\ \bibnamefont
  {Barato}}\ and\ \bibinfo {author} {\bibfnamefont {U.}~\bibnamefont
  {Seifert}},\ }\Doi {10.1103/PhysRevLett.114.158101} {\bibfield  {journal}
  {\bibinfo  {journal} {Phys. Rev. Lett.},\ }\textbf {\bibinfo {volume}
  {114}},\ \bibinfo {pages} {158101} (\bibinfo {year} {2015})}\BibitemShut
  {NoStop}%
\bibitem [{\citenamefont {Pietzonka}\ \emph
  {et~al.}(2016){\natexlab{a}}\citenamefont {Pietzonka}, \citenamefont
  {Barato},\ and\ \citenamefont {Seifert}}]{Pietzonka2016Affinity}%
  \BibitemOpen
  \bibfield  {author} {\bibinfo {author} {\bibfnamefont {P.}~\bibnamefont
  {Pietzonka}}, \bibinfo {author} {\bibfnamefont {A.~C.}\ \bibnamefont
  {Barato}}, \ and\ \bibinfo {author} {\bibfnamefont {U.}~\bibnamefont
  {Seifert}},\ }\Doi {10.1088/1751-8113/49/34/34LT01} {\bibfield  {journal}
  {\bibinfo  {journal} {J. Phys. A: Math. Theor.},\ }\textbf {\bibinfo {volume}
  {49}},\ \bibinfo {pages} {34LT01} (\bibinfo {year}
  {2016}{\natexlab{a}})}\BibitemShut {NoStop}%
\bibitem [{\citenamefont {Pietzonka}\ \emph
  {et~al.}(2016){\natexlab{b}}\citenamefont {Pietzonka}, \citenamefont
  {Barato},\ and\ \citenamefont {Seifert}}]{Pietzonka2016Universal}%
  \BibitemOpen
  \bibfield  {author} {\bibinfo {author} {\bibfnamefont {P.}~\bibnamefont
  {Pietzonka}}, \bibinfo {author} {\bibfnamefont {A.~C.}\ \bibnamefont
  {Barato}}, \ and\ \bibinfo {author} {\bibfnamefont {U.}~\bibnamefont
  {Seifert}},\ }\Doi {10.1103/PhysRevE.93.052145} {\bibfield  {journal}
  {\bibinfo  {journal} {Phys. Rev. E},\ }\textbf {\bibinfo {volume} {93}},\
  \bibinfo {pages} {052145} (\bibinfo {year} {2016}{\natexlab{b}})}\BibitemShut
  {NoStop}%
\bibitem [{\citenamefont {Gingrich}\ \emph {et~al.}(2016)\citenamefont
  {Gingrich}, \citenamefont {Horowitz}, \citenamefont {Perunov},\ and\
  \citenamefont {England}}]{Gingrich2016}%
  \BibitemOpen
  \bibfield  {author} {\bibinfo {author} {\bibfnamefont {T.}~\bibnamefont
  {Gingrich}}, \bibinfo {author} {\bibfnamefont {J.~M.}\ \bibnamefont
  {Horowitz}}, \bibinfo {author} {\bibfnamefont {N.}~\bibnamefont {Perunov}}, \
  and\ \bibinfo {author} {\bibfnamefont {J.~L.}\ \bibnamefont {England}},\
  }\href@noop {} {\bibfield  {journal} {\bibinfo  {journal} {Phys. Rev.
  Lett.},\ }\textbf {\bibinfo {volume} {116}},\ \bibinfo {pages} {120601}
  (\bibinfo {year} {2016})}\BibitemShut {NoStop}%
\bibitem [{\citenamefont {Gingrich}\ \emph {et~al.}(2017)\citenamefont
  {Gingrich}, \citenamefont {Rotskoff},\ and\ \citenamefont
  {Horowitz}}]{Gingrich2017}%
  \BibitemOpen
  \bibfield  {author} {\bibinfo {author} {\bibfnamefont {T.~R.}\ \bibnamefont
  {Gingrich}}, \bibinfo {author} {\bibfnamefont {G.~M.}\ \bibnamefont
  {Rotskoff}}, \ and\ \bibinfo {author} {\bibfnamefont {J.~M.}\ \bibnamefont
  {Horowitz}},\ }\Doi {10.1088/1751-8121/aa672f} {\bibfield  {journal}
  {\bibinfo  {journal} {J. Phys. A: Math. Theor.},\ }\textbf {\bibinfo {volume}
  {50}},\ \bibinfo {pages} {184004} (\bibinfo {year} {2017})}\BibitemShut
  {NoStop}%
\end{thebibliography}%

\end{document}